\documentclass[aps,prl,twocolumn,preprintnumbers,10pt,showpacs,nofootinbib]{revtex4-1}
\usepackage{amsmath,bm,amssymb}

\usepackage{graphicx}
\usepackage{psfrag}
\usepackage[caption=false]{subfig}

\newcommand\eps{\epsilon}

\def\cO{  {\cal O}  }

\newcommand{\cA}{{\cal A}}

\def\eps{\epsilon}
\def\cO{  {\cal O}  }
\newcommand{\bea}{\begin{eqnarray}}
\newcommand{\eea}{\end{eqnarray}}

\def\beq{\begin{equation}}   
\def\eeq{\end{equation}}
\def\bea{\begin{eqnarray}}  
\def\eea{\end{eqnarray}} 

\usepackage{slashed}

\def\Ell{{(L)}}
\def\Ellk{{(L,k)}}
\def\Elltwok{{(L,2k)}}
\def\Elltwokplus{{(L,2k+1)}}

\def \be  {\begin{equation}}
\def \ee  {\end{equation}}
\def \ba  {\begin{eqnarray}}
\def \ea  {\end{eqnarray}}
\def\la {\lambda}

\begin{document}

\preprint{CERN-TH-2016-174, MITP/16-084}

\title{Four-gluon scattering at three loops, infrared structure and Regge limit}

\author{J.\ M.\ Henn$^a$, B.\ Mistlberger$^b$}

\affiliation{
$^a$ 
PRISMA Cluster of Excellence, 
Johannes Gutenberg University, 55099 Mainz, Germany\\
$^b$ CERN, Geneva, Switzerland}


\begin{abstract}
We compute the three-loop four-gluon scattering amplitude in maximally supersymmetric Yang-Mills theory, including its full color dependence. 
Our result is the first complete computation of a non-planar four-particle scattering amplitude to three loops in four-dimensional gauge theory
and consequently provides highly non-trivial data for the study of non-planar scattering amplitudes.
We present the amplitude as a Laurent expansion in the dimensional regulator to finite order, with coefficients composed of harmonic poly-logarithms of uniform transcendental weight, and simple rational prefactors.
Our computation 
provides an independent check of a recent result for three-loop corrections to the soft anomalous dimension matrix that predicts the general infrared singularity structure of massless gauge theory scattering amplitudes.
Taking the Regge limit of our result, we determine the three-loop gluon Regge trajectory.
We also find agreement with 
very recent predictions 
for sub-leading logarithms.
\end{abstract}

\maketitle

\section{Introduction}

In the era of the Large Hadron Collider our understanding of the fundamental interactions of nature is probed at an unprecedented level. An incredible variety of observables is measured at rapidly increasing precision. 
To fully exploit the information collected from particle collisions the experimental progress has to be matched by an equally rapid theoretical development of our capabilities to predict the outcome of scattering experiments. 
Only the fruitful symbiosis of precise prediction and measurement can lead to an improved understanding of nature. 
Meeting these theoretical challenges pushes the boundaries of our understanding of the very structure of quantum field theory to its limits and is often accompanied by conceptual breakthroughs.

The maximally supersymmetric Yang-Mills theory serves as a useful testing laboratory for the development of novel methods for precise predictions and is used to explore hidden structures of quantum field theory.
In many aspects this theory can be 
thought of as a toy model for the Standard Model of particle physics, very much in analogy to the hydrogen atom with respect to quantum mechanics \cite{Caron-Huot:2014gia}.
Novel on-shell methods \cite{Bern:1994zx,Witten:2003nn,Britto:2005fq,ArkaniHamed:2010kv},
new techniques for working with transcendental functions \cite{Goncharov:2010jf}, 
and for computing them efficiently \cite{Henn:2013pwa}, were first developed in this theory.

An important catalyst for such advances is readily available analytic `data' for scattering amplitudes,
as such information makes it possible to test new approaches and to search for and uncover hidden simplicity. 
The large majority of known results for scattering amplitudes are available for the planar sector of the theory, which has a Yangian symmetry \cite{Drummond:2009fd} and is believed to be integrable. 
For example, the exact functional form of the four-particle amplitude in the planar limit is known to all loop orders \cite{Anastasiou:2003kj,Bern:2005iz,Drummond:2007au}. 
Moreover, more general scattering amplitudes can be computed using integrability methods, see e.g. \cite{Alday:2010vh,Basso:2014nra}.

The non-planar sector of scattering amplitudes is 
still
much less explored and analytic results are scarce in comparison.
Most studies to date are limited to loop integrands, uncovering that they exhibit several intriguing features that hint at 
more underlying structure, and perhaps a dual formulation of the theory \cite{ArkaniHamed:2009dn}.
A key concept is the analysis of leading singularities,
i.e. multidimensional residues. For example, it can be shown that loop integrands of maximally-helicity-violating amplitudes have a very simple leading singularity structure: any of their leading singularities is proportional to a tree-level amplitude \cite{ArkaniHamed:2010gh}.
Moreover, the integrand of these amplitudes can be written in terms of certain d-log forms \cite{ArkaniHamed:2012nw}.
This is closely related to a further observation, namely the absence of poles at infinity in the sense of \cite{Arkani-Hamed:2014via,Bern:2014kca}. 
In the planar case this follows from Yangian symmetry, but at the non-planar level this interesting property is not yet proven. 
The above properties may also hint at a connection to the amplituhedron interpretation of loop integrands \cite{Arkani-Hamed:2013jha,Bern:2015ple}.
Moreover, there is evidence that the color-kinematics duality found in \cite{Bern:2008qj} holds at loop level, see e.g. \cite{Bern:2012uf}.
The latter connects planar and non-planar terms in the integrand, and moreover relates loop integrands in Yang-Mills theory to their 
(super-)gravity counterparts. This may be significant for investigations into the possible finiteness of certain supergravity theories, see e.g. \cite{Bern:2008pv}.

It is important to ask how and if these intriguing loop-integrand properties propagate to the integrated answer.
One example where this connection has been made concrete, albeit at the conjectural level, has to do with the class of functions expected to appear.
Based on explicitly known results it is believed that scattering amplitudes
in ${\mathcal N}=4$ super Yang-Mills (SYM) are given by functions of uniform transcendental weight\footnote{For multiple polylogarithms, the weight is defined as the number of integrations. The dimensional regularization parameter is assigned weight $-1$ \cite{Henn:2013pwa}.}.
A conjecture due to \cite{ArkaniHamed:2010gh} links this property to integrands that can be written in terms of d-log forms \cite{ArkaniHamed:2012nw}, and hence have constant leading singularities.
We remark that these concepts have already proven to be very useful beyond maximally supersymmetric Yang-Mills theory \cite{Henn:2013pwa} in computing large classes of planar and non-planar loop integrals relevant for collider physics. 

On the other hand it is currently unknown what the precise implications 
of the other fascinating integrand properties mentioned above are for the integrated answer.
It goes without saying that explicit analytic results are very important for making progress 
in this direction.
To date, the only known non-planar amplitudes in the maximally super Yang-Mills theory are the two-loop four-point amplitudes
\cite{Bern:1997nh,Naculich:2008ys,Naculich:2013xa},
as well as a three-loop form factor that was computed via non-planar integrals \cite{Gehrmann:2011xn}.

In this work, we present the complete three-loop four-gluon amplitude in maximally supersymmetric Yang-Mills theory.
Our result is a new milestone in the advancement of perturbative quantum field theory as it represents the first complete
 computation of a non-planar four-particle scattering amplitude at three-loop order in four-dimensional perturbative quantum field theory. 
We investigate some of the remarkable properties of $\mathcal{N}=4$ SYM introduced above and study the infrared and high energy properties of our result.

The paper is organized as follows. 
We begin by explaining how we performed the calculation starting from several different representations for the integrands of the scattering amplitude. 
We review the color structure of the amplitude and explain how the result can be parametrized by a small number of independent components. 
Then, we discuss the structure of infrared divergences and confirm a recently obtained formula for contributions depending in a non-trivial way on four particles. Having removed the infrared divergences, we define a renormalized finite part.
Finally, we study the Regge limit of our result.
All our results are available in electronic form as ancillary files.

\section{Four-particle amplitudes}
The ${\mathcal N}=4$ super Yang-Mills four-gluon scattering amplitudes can be written perturbatively as 
\beq
\mathcal{A}(p_i;\epsilon)=  \mathcal{K} \sum\limits_{L=0}^\infty  \alpha^L\mathcal{A}^{(L)}(s,t;\epsilon)\,.
\eeq
We introduced the expansion parameter $\alpha = 4 e^{- \eps \gamma_{\rm E}} g^2/(4 \pi)^{2-\eps}$, where $\eps$ is the dimensional regularization parameter, with $D=4-2 \eps$.
Here the helicity structure was absorbed into a permutation invariant version of the tree-level amplitudes $\mathcal{K}$. It is explicitly given by 
\beq
\mathcal{K}  = 
- 2 s t g^2
\frac{\delta^{(8)}(Q)\delta^{(4)}(P)}{\langle 1 2 \rangle  \langle 2 3 \rangle  \langle 3 4 \rangle  \langle 4 1 \rangle }\,,
\eeq
cf.  \cite{Nair:1988bq} for details on the notation.  
Here we will only need that $\mathcal{A}^{(L)}$ depends on the Lorentz invariants $s=2 p_1 \cdot p_2$ and $t= 2 p_2 \cdot p_3$. We also have $u=2 p_1 \cdot p_3 = -s-t$.

The integrand for the three-loop four-particle scattering amplitude was investigated in a number of papers.
The main obstacle for computing the amplitudes were the complicated three-loop non-planar Feynman integrals.
Analytic results for the latter are now available \cite{Henn:2013fah,Henn:2013nsa,unpublished}. 
They were evaluated via the method of differential equations~\cite{Kotikov:1990kg,Gehrmann:1999as}, with a basis of integrals that have unit leading singularities \cite{Henn:2013pwa}.
We performed the necessary integral reductions to relate the basis chosen in \cite{unpublished} to those of references \cite{Bern:2008pv,Bern:2012uf,Bern:2014kca,Bern:2015ple}.
It is a highly non-trivial cross-check of the very different methods used to compute the integrand that we obtained the same result for each representation. 

We present our analytic result for the amplitude, computed in the dimensional reduction scheme, as a Laurent expansion in the dimensional 
regulator, to order $\eps^0$. 
The coefficient of each power of the dimensional regulator is expressed in terms of harmonic polylogarithms \cite{Remiddi:1999ew} and rational prefactors.
The answer has two interesting features that we would like to highlight before discussing the result in more detail.
First, we show that, to all orders in $\eps$, the amplitude has uniform transcendental weight.
Second, the only rational structures appearing are $1/s/t, 1/s/u$ and $1/t/u$, corresponding to 
different tree-level channels. 
This confirms, in a highly non-trivial case, expectations based on the properties of leading singularities \cite{ArkaniHamed:2010gh,Bern:2014kca}. 
We also wish to emphasize that although the latter were only analyzed in four dimensions, we find the remarkable fact that the uniform weight property is true to all orders in the $\eps$ expansion.

\section{Color decomposition}

The amplitude $\mathcal{A}$ is a tensor in color space. 
It can be decomposed to all orders in terms of traces of fundamental color generators of $SU(N_c)$. 
We abbreviate 
\beq
{\rm tr}(T^{a_1}T^{a_2}T^{a_3}T^{a_4})={\rm tr}(1234)\,,
\eeq
and normalize the fundamental generators according to ${\rm tr}(T^{a} T^{b}) = \delta^{ab}/2$.
At any order in the coupling constant, the amplitude can be expressed in terms of the following six single and double trace color structures,
\bea
C_{1}&=&{\rm tr}(1234 )+{\rm tr}(1432 )\,,\hspace{1cm} C_4={\rm tr}(12){\rm tr}(34) \,, \nonumber\\
C_{2}&=&{\rm tr}(1243 )+{\rm tr}(1342 )\,,\hspace{1cm} C_5={\rm tr}(13){\rm tr}(24) \,, \nonumber\\
C_{3}&=&{\rm tr}(1423 )+{\rm tr}(1324 )\,,\hspace{1cm} C_6={\rm tr}(14){\rm tr}(23) \,.\nonumber
\eea
Following \cite{Naculich:2011ep}, we further decompose the amplitude in powers of $N_c$,
\bea
\cA^\Ell &=&
\sum_{\la = 1}^3 
\left( \sum_{k=0}^{\lfloor \frac{L}{2}  \rfloor} N_c^{L-2k} A^\Elltwok_\la \right) C_\la \nonumber\\
&&+ \sum_{\la = 4}^6 
\left( \sum_{k=0}^{\lfloor \frac{L-1}{2}  \rfloor}N_c^{L-2k-1} A^\Elltwokplus_\la\right) C_\la \,,
\label{decomp}
\eea
where  $A^{(L,0)}_\la$ are leading-order-in-$N_c$ (planar) amplitudes,
and  $A^\Ellk_\la$, $k = 1, \cdots, L$,  are subleading,
yielding $(3L+3)$ color-ordered amplitudes at $L$ loops.

Some terms in eq. (\ref{decomp}) are related by group theory identities, such as the $U(1)$ decoupling relation.  For example, at one loop \cite{Bern:1990ux}, 
\be
A^{(1,1)}_4 = A^{(1,1)}_5 =A^{(1,1)}_6 = 2  \sum_{\lambda=1}^{3} A^{(1,0)}_\lambda \,.
\ee
Taking into account similar relations \cite{Naculich:2011ep} allows one to determine $A^{(2,1)}$ and $A^{(3,3)}$
in terms of the other components.
This means that, up to three loops, the amplitudes
can be expressed in terms of a small number of functions. 
Apart from the leading color terms $A^{(L,0)}_\la$,
one only needs the following components:
$A^{(2,2)}_1$  at two loops, and $A^{(3,2)}_1$ and $A^{(3,1)}_4$ at three loops.
All other terms can be obtained either by group theory relations, or by symmetry.
It is interesting to note that the remaining components still satisfy constraints coming from group theory, e.g. 
$\sum_{\lambda=1}^{3} A^{(2,2)}_\lambda = 0$.
$A^{(2,2)}_1$ was determined in ref. \cite{Naculich:2013xa}. 
In the present paper we compute the new components $A^{(3,2)}_1$ and $A^{(3,1)}_4$ at the three loop order.

\section{Infrared divergence structure}

The structure of infrared divergences of massless scattering amplitudes is well understood.
They can be mapped to ultraviolet divergences of Wilson loops, with the latter being controlled by renormalization group equations, see e.g. \cite{Korchemsky:1985xj, Korchemsky:1991zp}.
An amplitude for the scattering of massless, $SU(N_c)$ color charged fields 
in dimensional regularisation can be written as 
\beq
\mathcal{A}(p_i ,\epsilon)=\bold{Z}(p_i ,\epsilon) \mathcal{A}^f(p_i ,\epsilon)\,,
\eeq
where $\mathcal{A}^f(p_i )$ represents a finite hard amplitude. 
To indicate an operator in color space we use bold letters.
The factor $\bold Z(p_i ,\epsilon)$ contains all infrared divergences. 
It is given by the exponential
\beq\label{Zexp}
\bold{Z}(p_i,\epsilon) =\mathcal{P} exp\left\{-\frac{1}{2}\int_0^{\mu^2} \frac{d\mu^2}{\mu^2} \bold \Gamma(p_i,\mu^2,\alpha(\mu^2)) \right\}\,,
\eeq
where $\alpha(\mu^2)$ is the renormalised coupling constant, 
and ${\bf \Gamma}$ is the soft anomalous dimension.

In ${\mathcal N}=4$ super Yang-Mills, the renormalization of the coupling is trivial, and hence the integral in the exponent of eq. (\ref{Zexp}) can be carried out explicitly.
For four-gluon scattering one obtains
\bea
\label{eq:expformula}
 \frac{1}{4}\sum\limits_{L=1}^\infty
 \alpha^L 
 \left[\frac{\gamma_c^{(L)}}{L^2 \epsilon^2} \bold{D_0}-\frac{\gamma_c^{(L)} }{L \epsilon} \bold{D}+ \frac{4}{L \epsilon}   \gamma_J^{(L)} \bold{\mathbb{I}} + \frac{1}{L \eps} \bold{\Delta}^{(L)} \right] \,.
\eea
Here $\gamma_c$ is the cusp anomalous dimension and $\gamma_J$ is the collinear anomalous dimensions (associated with the external gluons). To three loops~\cite{Vogt:2004mw,Kotikov:2004er,Gehrmann:2010ue}, they read
\bea \label{eqcusp}
\gamma_c&=&\alpha -\frac{1}{2} \alpha^2 N_c \zeta_2 +\frac{11}{8} \alpha^3N_c^2 \zeta_4+\mathcal{O}\left(\alpha^4\right),\\
\gamma_J&=&\frac{1}{4} \alpha^2 N_c^2 \zeta_3-\frac{1}{2} \alpha^3 N_c^3 \left(\frac{5}{6}\zeta_2\zeta_3+\zeta_5\right)+\mathcal{O}\left(\alpha^4\right).
\eea
The first two color operators in eq. (\ref{eq:expformula}) correspond to dipole terms, i.e. they 
depend only pairwise on the incoming particles. They are given by
\bea
 \bold{D}_0= \sum\limits_{i
 \neq
  j}\bold{T}_i\cdot\bold{T}_j , \;\;
\bold{D}=\sum\limits_{i
\neq
 j}\bold{T}_i \cdot\bold{T}_j \log\left(\frac{-s_{ij}}{\mu^2}\right),
\eea
with $s_{ij}=2 p_i\cdot p_j$. 
The color operators act according to  $\bold{T}_1^{a_5} T^{a_1}  =- i f^{a_5 a_1 a_6} T^{a_6} $.

Up to two loops \cite{Catani:1998bh,Sterman:2002qn}, the soft anomalous dimension is given by a dipole formula, 
$\bold{\Delta}^{(1)}=\bold{\Delta}^{(2)}=\bold{0}$.
Three loop corrections to the latter are universal in any gauge theory, as the matter dependent terms cancel \cite{Dixon:2009gx}. 
They can be split into a contributions connecting three and four color charged external fields, 
and we refer to the latter as  $\bold{\Delta}_3^{(3)}$ and $\bold{\Delta}_4^{(3)}$, respectively. 
These corrections were obtained recently in ref.~\cite{Almelid:2015jia} for the case of $n$-particle scattering.
Restricting the general formula to the case of four-particle scattering we find
\bea\label{quadrupoleoperator}
\bold{\Delta}_4^{(3)} &=&  \frac{1}{4} \, f_{abe}f_{cde} \Big[ 
 {\rm \bf T}_1^a  {\rm \bf T}_2^b   {\rm \bf T}_3^c {\rm \bf T}_4^d   \, \mathcal{S}(x) 
   \nonumber \\&& 
  +{\rm \bf T}_4^a   {\rm \bf T}_1^b  {\rm \bf T}_2^c    {\rm \bf T}_3^d \, \mathcal{S}(1/x) \Big], \\ 
\bold{\Delta}_3^{(3)}&=&-  C \,f_{abe}f_{cde}  \sum_{\substack{{i=1 \ldots 4}\\{1\leq j<k\leq 4}\\ j,k\neq i}}\left\{{\rm \bf T}_i^a,  {\rm \bf T}_i^d\right\}   {\rm \bf T}_j^b {\rm \bf T}_k^c \,.  
\eea
Here $C=\zeta_5+2\zeta_3\zeta_2$, and $\mathcal{S}(x)$ is given by
\bea \label{softfunction}
\mathcal{S}(x) &=& \\
&& \hspace{-0.9cm}    2 H_{-3,-2}+2 H_{-2,-3}-2 H_{-3,-1,-1}+2 H_{-3,-1,0} \nonumber\\ && \hspace{-0.9cm} 
  -2 H_{-2,-2,-1}    +2 H_{-2,-2,0}-2
   H_{-2,-1,-2}-H_{-1,-2,-2} \nonumber\\ && \hspace{-0.9cm} 
-H_{-1,-1,-3}+4 H_{-2,-1,-1,-1}  -2
   H_{-2,-1,-1,0} \nonumber\\ && \hspace{-0.9cm} 
   -H_{-1,-2,-1,0}-H_{-1,-1,-2,0}    +\zeta _3 H_{-1,-1}
  +4 \zeta _3 \zeta _2-\zeta _5   \nonumber\\ && \hspace{-0.9cm} 
+ \zeta _2 ( 6  H_{-3}   -10 H_{-2,-1}+6  H_{-2,0}- H_{-1,-2} - H_{-1,-1,0}) \nonumber\\ && \hspace{-0.9cm} 
   + i \pi \Big[
2 H_{-3,-1}-2 H_{-3,0}+2 H_{-2,-2}-4 H_{-2,-1,-1} \nonumber\\ && \hspace{-0.9cm} \phantom{ + i \pi}
 +2 H_{-2,-1,0}-2
   H_{-2,0,0}+H_{-1,-2,0}+H_{-1,-1,0,0}\nonumber\\ && \hspace{-0.9cm}  \phantom{ + i \pi}
+   \zeta_2 ( 3   H_{-1,-1}-4  H_{-2}) -\zeta _3 H_{-1} \Big] \,.\nonumber
\eea
Here $H$ are harmonic polylogarithms, and we have suppressed
the argument $x=t/s$.
The imaginary part in eq. (\ref{softfunction}) deserves a comment. 
It arises when analytically continuing the result of \cite{Almelid:2015jia}, which is written for $s_{ij}<0$, where  the
the soft anomalous dimension matrix is real-valued, to four-particle kinematics with $u=-s-t > 0$.
%
Given eq.~\eqref{eq:expformula} we can obtain explicitly all the entire infrared pole structure of our four point amplitudes. 
Comparing the above predictions of the infrared singularities for the four-particle amplitude with our result we find perfect agreement.

\section{Results for the amplitude}

Using the knowledge about the universal infrared structure of our amplitude we can inspect in detail the finite remainder after subtracting the divergences.
\begin{align}\label{definitonfinite}
\mathcal{H} = \lim_{\eps \to 0} \mathcal{A}^f \,. 
\end{align}
We will use the color decomposition as before, and define $H$ equivalently to $A$ in eq.~\eqref{decomp} but for infrared finite amplitudes.
In the planar limit, the finite part is given by a remarkably simple formula that was conjectured in \cite{Anastasiou:2003kj,Bern:2005iz} and proven in \cite{Drummond:2007au}, 
\bea
\sum_{L} \alpha^L H^{(L,0)}_1  &=& H^{(0,0)}_1  \exp \Big\{ - \frac{ N_{c} \gamma_{\rm c}(\alpha)}{2}  \log \frac{-s}{\mu^2} \log \frac{-t}{\mu^2}  \nonumber \\
&& \hspace{-1cm}-\frac{ \gamma_J(\alpha)}{2}  \left[   \log \frac{-s}{\mu^2} + \log \frac{-t}{\mu^2} \right]   + C(\alpha)     \Big\} \,,
\eea
to all loop orders. The coupling dependence enters only through kinematic-independent constants.
%

In the present paper we evaluated the non-leading coefficients in the large $N_{c}$ expansion.
The two-loop function $H^{(2,2)}_1$ is expressed in terms of weight four harmonic polylogarithms,
while the two new functions $H^{(3,2)}_1$ and $H^{(3,1)}_4$ at three loops are given by 
uniform weight six harmonic polylogarithms. 
Their expressions, as well as those for the unrenormalized amplitude, can be found in an ancillary file.

\section{Regge limit}

Our analytic three-loop result gives us the opportunity to investigate the Regge behaviour of the amplitude, i.e. the kinematic regime of $s \gg t$. In the following, we work at leading power in $t/s$, and set $\mu^2 = -t$ for simplicity.


We found it useful to work in a color basis corresponding to irreducible $SU(N)$ representations in the tensor product of 
the representations of two gluons in the $t$-channel, following \cite{Kidonakis:1998nf,Korchemsky:1993hr,Dokshitzer:2005ek,DelDuca:2014cya}. 
Labelling the channels by their dimensions for $N_{c}=3$ (but continuing to work with general $N_c$), they are ${\bf 1}, {\bf 8}_{s}, {\bf 8}_a, {\bf 10}+\overline{{\bf 10}}, {\bf 27}, {\bf 0}$. 

In the octet channel ${\bf 8}_a$, we find that the amplitude behaves as 
$\mathcal{A}_{{\bf 8}_{a}}  \sim  s^{w_{{\bf 8}_{a}}}$, 
with the gluon Regge trajectory at three loops
\begin{align}
w_{{\bf 8}_{a}} |_{\alpha^3}=&  N^3_{c} \left[ \frac{11 \zeta_4}{48} \frac{1}{\eps} + \frac{5 }{24} \zeta_2 \zeta_{3} +  \frac{1}{4}  \zeta_{5}+\cO(\eps) \right] \nonumber\\ & 
\hspace{-0.1cm}+ N_{c} \left[ \frac{\zeta_2}{4} \frac{1}{\eps^3} - \frac{15 \zeta_4}{16} \frac{1}{\eps} - \frac{77}{4} \zeta_2 \zeta_{3}+\cO(\eps) \right] \,.
\end{align}
The leading color term is well known \cite{Naculich:2007ub,DelDuca:2008jg}. 
At subleading color, the infrared divergent terms were discussed in \cite{DelDuca:2014cya}; the finite term is new.
As far as other channels are concerned, we find perfect agreement with the result of \cite{DelDuca:2011ae,Caron-Huot:2013fea} for next-to-leading logarithms. Moreover, we successfully compared against very recent results at the next-to-next-to leading logarithmic level for the ${\bf 10}+\overline{{\bf 10}}$ channel \cite{simontoappear}.

Let us discuss in more detail the Regge limit of the finite part ${\mathcal H}$, 
as defined in eq. (\ref{definitonfinite}).
We write it in terms of color operators ${\bf T}=({\bf  T}_2+{\bf T}_3)^2$ and ${\bf S}=({\bf  T}_1+{\bf T}_2)^2$ acting 
on the tree-level amplitude ${\mathcal H}^{(0)}$.
All the logarithmically enhanced terms are given by
\begin{align}
{\mathcal H} =& \sum_{k,q}  \alpha^k  \left(\log \frac{s}{t} \right)^q {\bf O}_{k,q}  {\mathcal H}^{(0)} + {\cO(1)} \,,
\end{align}
where the sum includes the following operators, 
\begin{align}\label{operatorsO}
{\bf O}_{2,1}= & -\frac{1}{8}   \zeta_3 {\bf T}^2 \,, \\
{\bf O}_{3,2} =&  i \pi  \frac{11  }{24} \zeta_3 [[{\bf S},{\bf T}],{\bf T}]  \,, \\
{\bf O}_{3,1} =&   i \pi  \frac{1  }{16} \zeta_4    \left(  3   [ {\bf S},{\bf T} ] {\bf T} 
 +    58  [[{\bf S},{\bf T}],{\bf T}] \right)   \nonumber\\ &  
+   \frac{11}{6} \zeta_2 \zeta_3  \left( 3  [ {\bf S},{\bf T} ] {\bf T}  + 2  [[{\bf S},{\bf T}],{\bf T}] -  [{\bf S}^2,{\bf T}]
 \right)     
\nonumber\\ &
+   \left(   \frac{1}{4}  \zeta_5  -\frac{1}{24} \zeta_2 \zeta_3  \right)   {\bf T}^3 - 4 \zeta_2 \zeta_3  {\bf T} \,.
\end{align}
It is interesting to note that there is no contribution to the ${\mathbf 8}_{s}$ channel; the only term contributing to ${\bf 10}+\overline{{\bf 10}}$ is $ [{\bf S}^2,{\bf T}]$, and the contribution to ${\mathbf 8}_{a}$ comes from the ${\bf T},{\bf T}^2$, and ${\bf T}^3$ terms only.

\section{Conclusion}

We have computed for the first time a complete three-loop four-gluon scattering amplitude including all sub-leading color contributions.
This allowed us to verify a recent result for the universal structure of infrared divergences at the three-loop order. 
The four particle scattering amplitude in $\mathcal{N}=4$ SYM is a uniform weight combination of harmonic polylogarithms, to any order in the $\eps$ expansion.
We analyzed the Regge limit and determined the three-loop Regge trajectory. 
Our full results are given in ancillary files, and provide a non-trivial new data point
in the systematic exploration of properties of non-planar scattering amplitudes.

\section{Acknowledgments}

We thank C.~Duhr and G.~Korchemsky for helpful discussions, and S.~Caron-Huot for correspondence on \cite{simontoappear}.
J.M.H. is supported in part by a GFK fellowship and by
the PRISMA cluster of excellence at Mainz university.
B.M. is supported by the European Commission through the ``HICCUP'' ERC grant.



\begin{thebibliography}{99}

\bibitem{Caron-Huot:2014gia} 
  S.~Caron-Huot and J.~M.~Henn,
  Phys.\ Rev.\ Lett.\  {\bf 113}, no. 16, 161601 (2014).


\bibitem{Bern:1994zx} 
  Z.~Bern, L.~J.~Dixon, D.~C.~Dunbar and D.~A.~Kosower,
  Nucl.\ Phys.\ B {\bf 425}, 217 (1994).

\bibitem{Witten:2003nn} 
  E.~Witten,
  Commun.\ Math.\ Phys.\  {\bf 252}, 189 (2004).

\bibitem{Britto:2005fq} 
  R.~Britto, F.~Cachazo, B.~Feng and E.~Witten,
  Phys.\ Rev.\ Lett.\  {\bf 94}, 181602 (2005).

\bibitem{ArkaniHamed:2010kv} 
  N.~Arkani-Hamed, J.~L.~Bourjaily, F.~Cachazo, S.~Caron-Huot and J.~Trnka,
  JHEP {\bf 1101}, 041 (2011).


\bibitem{Goncharov:2010jf} 
  A.~B.~Goncharov, M.~Spradlin, C.~Vergu and A.~Volovich,
  Phys.\ Rev.\ Lett.\  {\bf 105}, 151605 (2010).
  
\bibitem{Henn:2013pwa} 
  J.~M.~Henn,
  Phys.\ Rev.\ Lett.\  {\bf 110}, 251601 (2013).
  
  

\bibitem{Drummond:2009fd} 
  J.~M.~Drummond, J.~M.~Henn and J.~Plefka,
  JHEP {\bf 0905}, 046 (2009).
  

 
\bibitem{Anastasiou:2003kj} 
  C.~Anastasiou, Z.~Bern, L.~J.~Dixon and D.~A.~Kosower,
  Phys.\ Rev.\ Lett.\  {\bf 91}, 251602 (2003).
  
    
\bibitem{Bern:2005iz} 
  Z.~Bern, L.~J.~Dixon and V.~A.~Smirnov,
  Phys.\ Rev.\ D {\bf 72}, 085001 (2005).
  
\bibitem{Drummond:2007au} 
  J.~M.~Drummond, J.~Henn, G.~P.~Korchemsky and E.~Sokatchev,
  Nucl.\ Phys.\ B {\bf 826}, 337 (2010).
  
  
\bibitem{Basso:2014nra} 
  B.~Basso, A.~Sever and P.~Vieira,
  JHEP {\bf 1409}, 149 (2014).
 
\bibitem{Alday:2010vh} 
  L.~F.~Alday, J.~Maldacena, A.~Sever and P.~Vieira,
  J.\ Phys.\ A {\bf 43}, 485401 (2010).
  
  
\bibitem{ArkaniHamed:2009dn} 
  N.~Arkani-Hamed, F.~Cachazo, C.~Cheung and J.~Kaplan,
  JHEP {\bf 1003}, 020 (2010).
  
  
  
 
  
\bibitem{ArkaniHamed:2010gh} 
  N.~Arkani-Hamed, J.~L.~Bourjaily, F.~Cachazo and J.~Trnka,
  JHEP {\bf 1206}, 125 (2012).
 













  
\bibitem{ArkaniHamed:2012nw} 
  N.~Arkani-Hamed, J.~L.~Bourjaily, F.~Cachazo, A.~B.~Goncharov, A.~Postnikov and J.~Trnka,
  arXiv:1212.5605 [hep-th].

  
  
\bibitem{Arkani-Hamed:2014via} 
  N.~Arkani-Hamed, J.~L.~Bourjaily, F.~Cachazo and J.~Trnka,
  Phys.\ Rev.\ Lett.\  {\bf 113}, no. 26, 261603 (2014).
  
\bibitem{Bern:2014kca} 
  Z.~Bern, E.~Herrmann, S.~Litsey, J.~Stankowicz and J.~Trnka,
  JHEP {\bf 1506}, 202 (2015).
  

\bibitem{Arkani-Hamed:2013jha} 
  N.~Arkani-Hamed and J.~Trnka,
  JHEP {\bf 1410}, 030 (2014).
 
 
 
\bibitem{Bern:2015ple} 
  Z.~Bern, E.~Herrmann, S.~Litsey, J.~Stankowicz and J.~Trnka,
  JHEP {\bf 1606}, 098 (2016).
  
  
\bibitem{Bern:2008qj} 
  Z.~Bern, J.~J.~M.~Carrasco and H.~Johansson,
  Phys.\ Rev.\ D {\bf 78}, 085011 (2008).


 
  
\bibitem{Bern:2012uf} 
  Z.~Bern, J.~J.~M.~Carrasco, L.~J.~Dixon, H.~Johansson and R.~Roiban,
  Phys.\ Rev.\ D {\bf 85}, 105014 (2012).

\bibitem{Bern:2008pv} 
  Z.~Bern, J.~J.~M.~Carrasco, L.~J.~Dixon, H.~Johansson and R.~Roiban,
  Phys.\ Rev.\ D {\bf 78}, 105019 (2008).



\bibitem{Bern:1997nh} 
  Z.~Bern, J.~S.~Rozowsky and B.~Yan,
  Phys.\ Lett.\ B {\bf 401}, 273 (1997).



\bibitem{Naculich:2008ys} 
  S.~G.~Naculich, H.~Nastase and H.~J.~Schnitzer,
  JHEP {\bf 0811}, 018 (2008).
  
  
  
\bibitem{Naculich:2013xa} 
  S.~G.~Naculich, H.~Nastase and H.~J.~Schnitzer,
  JHEP {\bf 1304}, 114 (2013).
  
  
  
\bibitem{Gehrmann:2011xn} 
  T.~Gehrmann, J.~M.~Henn and T.~Huber,
  JHEP {\bf 1203}, 101 (2012).




\bibitem{Nair:1988bq} 
  V.~P.~Nair,
  Phys.\ Lett.\ B {\bf 214}, 215 (1988).



  
\bibitem{Henn:2013nsa} 
  J.~M.~Henn, A.~V.~Smirnov and V.~A.~Smirnov,
  JHEP {\bf 1403}, 088 (2014).

\bibitem{Henn:2013fah} 
  J.~M.~Henn, A.~V.~Smirnov and V.~A.~Smirnov,
  JHEP {\bf 1307}, 128 (2013).


\bibitem{unpublished} 
  J.~M.~Henn, B.~Mistlberger and V.~A.~Smirnov,
 to appear.
 
\bibitem{Kotikov:1990kg}
  A.~V.~Kotikov,
  Phys.\ Lett.\ B {\bf 254} (1991) 158.

\bibitem{Gehrmann:1999as}
  T.~Gehrmann and E.~Remiddi,
  Nucl.\ Phys.\ B {\bf 580} (2000) 485.


\bibitem{Remiddi:1999ew} 
  E.~Remiddi and J.~A.~M.~Vermaseren,
  Int.\ J.\ Mod.\ Phys.\ A {\bf 15}, 725 (2000).
  
  
 
 
\bibitem{Naculich:2011ep} 
  S.~G.~Naculich,
  Phys.\ Lett.\ B {\bf 707}, 191 (2012).
  
  
\bibitem{Bern:1990ux} 
  Z.~Bern and D.~A.~Kosower,
  Nucl.\ Phys.\ B {\bf 362}, 389 (1991).




\bibitem{Korchemsky:1985xj} 
  G.~P.~Korchemsky and A.~V.~Radyushkin,
  Phys.\ Lett.\ B {\bf 171}, 459 (1986).

\bibitem{Korchemsky:1991zp} 
  G.~P.~Korchemsky and A.~V.~Radyushkin,
  Phys.\ Lett.\ B {\bf 279}, 359 (1992).


\bibitem{Kotikov:2004er}
  A.~V.~Kotikov, L.~N.~Lipatov, A.~I.~Onishchenko and V.~N.~Velizhanin,
  Phys.\ Lett.\ B {\bf 595} (2004) 521.
  
\bibitem{Vogt:2004mw}
  A.~Vogt, S.~Moch and J.~A.~M.~Vermaseren,
  Nucl.\ Phys.\ B {\bf 691} (2004) 129.
  
\bibitem{Gehrmann:2010ue}
  T.~Gehrmann, E.~W.~N.~Glover, T.~Huber, N.~Ikizlerli and C.~Studerus,
  JHEP {\bf 1006} (2010) 094.



\bibitem{Catani:1998bh} 
  S.~Catani,
  Phys.\ Lett.\ B {\bf 427}, 161 (1998).

\bibitem{Sterman:2002qn} 
  G.~F.~Sterman and M.~E.~Tejeda-Yeomans,
  Phys.\ Lett.\ B {\bf 552}, 48 (2003).
  
  
\bibitem{Almelid:2015jia} 
  O.~Almelid, C.~Duhr and E.~Gardi,
  arXiv:1507.00047 [hep-ph].

\bibitem{Dixon:2009gx} 
  L.~J.~Dixon,
  Phys.\ Rev.\ D {\bf 79}, 091501 (2009).









\bibitem{Kidonakis:1998nf} 
  N.~Kidonakis, G.~Oderda and G.~F.~Sterman,
  Nucl.\ Phys.\ B {\bf 531}, 365 (1998).
  
\bibitem{Korchemsky:1993hr} 
  G.~P.~Korchemsky,
  Phys.\ Lett.\ B {\bf 325}, 459 (1994).

  
  
\bibitem{Dokshitzer:2005ek} 
  Y.~L.~Dokshitzer and G.~Marchesini,
  Phys.\ Lett.\ B {\bf 631}, 118 (2005).

\bibitem{DelDuca:2014cya}
  V.~Del Duca, G.~Falcioni, L.~Magnea and L.~Vernazza,
  JHEP {\bf 1502} (2015) 029.
  
\bibitem{Naculich:2007ub} 
  S.~G.~Naculich and H.~J.~Schnitzer,
  Nucl.\ Phys.\ B {\bf 794}, 189 (2008).

\bibitem{DelDuca:2008jg} 
  V.~Del Duca, C.~Duhr and E.~W.~N.~Glover,
  JHEP {\bf 0812}, 097 (2008).
  
\bibitem{DelDuca:2011ae}
  V.~Del Duca, C.~Duhr, E.~Gardi, L.~Magnea and C.~D.~White,
  JHEP {\bf 1112} (2011) 021
 
 
\bibitem{Caron-Huot:2013fea} 
  S.~Caron-Huot,
  JHEP {\bf 1505}, 093 (2015).

  

\bibitem{simontoappear} 
  S.~Caron-Huot, E.~Gardi, L.~Vernazza,
  to appear.







 \end{thebibliography}
\end{document}